\documentclass[%
 reprint,
 superscriptaddress,
 showpacs,preprintnumbers,
 amsmath,amssymb,
 aps,
 prl,
 longbibliography,
lengthcheck,%
]{revtex4-2}

\usepackage{graphicx}% Include figure files
\usepackage{dcolumn}% Align table columns on decimal point
\usepackage{bm}% bold math
\usepackage{hyperref}% add hypertext capabilities
\usepackage{color}
\usepackage{ulem}

\begin{document}

%\preprint{APS/123-QED}

\title{
Skew scattering by magnetic monopoles and anomalous Hall effect in spin-orbit coupled systems
}

\author{Jun Mochida}
\affiliation{
Department of Physics, Tokyo Institute of Technology, Meguro, Tokyo, 152-8551, Japan
}
\affiliation{Department of Physics, University of Tokyo, 7-3-1 Hongo, Bunkyo-ku, Tokyo 113-0033, Japan
}

\author{Hiroaki Ishizuka}
\affiliation{
Department of Physics, Tokyo Institute of Technology, Meguro, Tokyo, 152-8551, Japan
}
\email{ishizuka@phys.titech.ac.jp}

\date{\today}

\begin{abstract}
Magnetic textures like skyrmions and domain walls coupled to itinerant electrons give rise to rich transport phenomena such as anomalous Hall effect and nonreciprocal current.
An interesting case is when the transport coefficient is related to the global (or topological) property of the magnetic texture, e.g., skyrmion and domain wall numbers.
Such phenomena are also interesting from applications, which the transport phenomena are potential probe for electrically detecting magnetic textures in nano-scale devices.
Here, we show that an anomalous Hall effect proportional to the net magnetic monopole charge occurs from skew scattering when the magnetic texture couples to itinerant electrons in a non-centrosymmetric system with spin-orbit interaction.
This mechanism gives rise to a finite anomalous Hall effect in a ferromagnetic domain wall whose spins rotate in the $xy$ plane, despite no out-of-plane magnetic moment.
We also discuss the relation between the magnetic texture contributing to the anomalous Hall effect and the crystal symmetry.
The results demonstrate rich features arising from the interplay of spin-orbit interaction and magnetic textures and their potential for detecting various magnetic textures in nanoscale devices.
\end{abstract}

%\pacs{
%}% PACS, the Physics and Astronomy
% Classification Scheme.

\maketitle

%%%%   Introduction   %%%%%%%%%%%%%%%%%%%%%%%%%%%%%%%%%%%%%%%%%%%%%%%%
%\section{Introduction}
Noncollinear magnetic textures give rise to numbers of novel phenomena, such as anomalous~\cite{Ye1999,Ohgushi2000,Chen2014} and spin~\cite{Ishizuka2013,Ishizuka2021} Hall effects, multiferroics~\cite{Katsura2005,Bulaevskii2008}, and electrical magnetochiral effect~\cite{Ishizuka2020}.
These phenomena are often related to scalar and vector spin chiralities defined by $\bm S_i\cdot\bm S_j\times\bm S_k$ and $\bm S_i\times\bm S_j$, respectively.
These phenomena were also discussed experimentally. For instance, the anomalous Hall effect (AHE) were studied in transition metal magnets, such as in materials with non-coplanar magnetic order~\cite{Taguchi2001} and magnetic skyrmions~\cite{Neubauer2009,Kurumaji2019}, and the electrical magnetochiral effect in helical magnets~\cite{Yokouchi2017,Aoki2019}. 

Among various examples, a particularly interesting example is the continuum limit, in which the scalar spin chirality is related to the skyrmion number~\cite{Ye1999}.
In the limit, the Hall conductivity is related to the skyrmion density, as discussed in the perturbation theory~\cite{Onoda2004} and the skew scattering argument~\cite{Ishizuka2018a}.
The relation to a topological quantity implies the robustness of the AHE against fluctuations in the spin texture.
Such robustness is particularly attractive for applications as it is a potential robust probe for detecting a skyrmion in nano-scale devices, such as race-track memory~\cite{Fert2013,Tomasello2014,Maccariello2018}.
However, such a relation between the transport coefficient and a quantity characterizing the magnetic texture is rare.

In this work, we study a relation between AHE and the magnetic monopole charge in two-dimensional (2d) systems with spin-orbit interaction (SOI).
As a demonstration, we study the anomalous Hall effect in a 2d electron system with Rashba interaction focusing on the skew scattering by magnetic moments.
Our calculation shows that, when the magnetic moments lie in the $xy$ plane, the Hall conductivity is proportional to the magnetic monopole charge characterized by the divergence of magnetic moments.
We also discuss that the existence of the monopole term is understandable based on the crystal symmetry, which we demonstrate by comparing Rashba and Dresselhaus models.
The results unveil a nontrivial relation between the magnetic charge and transport phenomena, a relation unique to systems with SOI.

AHE generally occurs in magnetic metals with SOI~\cite{Nagaosa2010}.
Pioneering works focus on the AHE in ferromagnets, where the anomalous Hall conductivity follows the magnetization curve~\cite{Hall1881}.
Theoretically, the mechanism of AHE is broadly classified into two groups: the intrinsic mechanism related to Berry curvature~\cite{Karplus1954} and the extrinsic mechanism involving impurity scattering~\cite{Smit1955,Berger1970}.
For the latter, the effect of SOI in the bulk band~\cite{Adams1959,Ishizuka2017} and magnetic impurity scattering~\cite{Kondo1962,Fert1987,Yamada1993} were also discussed in addition to the SOI of impurities.
In addition to the ferromagnets, recent studies find that non-coplanar magnetic states induce the AHE~\cite{Ye1999,Ohgushi2000,Onoda2004}.
The AHE by non-coplanar magnetic order raised interesting questions on the interplay of non-colinear magnetic states and the SOI, which gives rise to rich behaviors in AHE~\cite{Chen2014,Zhang2020,Lux2020,Yokoyama2020,Yamaguchi2021}.
Many works along this direction focus on the intrinsic mechanism, namely the AHE by spin Berry phase.

On the other hand, recently, a skew scattering mechanism by magnetic texture~\cite{Ishizuka2018a} has been proposed~\cite{Kanazawa2011}.
This mechanism potentially gives rise to a larger Hall effect compared to the intrinsic counterpart~\cite{Ishizuka2021}, as observed in recent experiments~\cite{Yang2020,Fujishiro2021,Uchida2021}.
However, these works discuss the skew scattering in electrons without SOI.
We here explore how the SOI affects the AHE by skew scattering, discussing the rich features arising from SOI and their relation to the crystal symmetry. 

\begin{figure}
  \includegraphics[width=\linewidth]{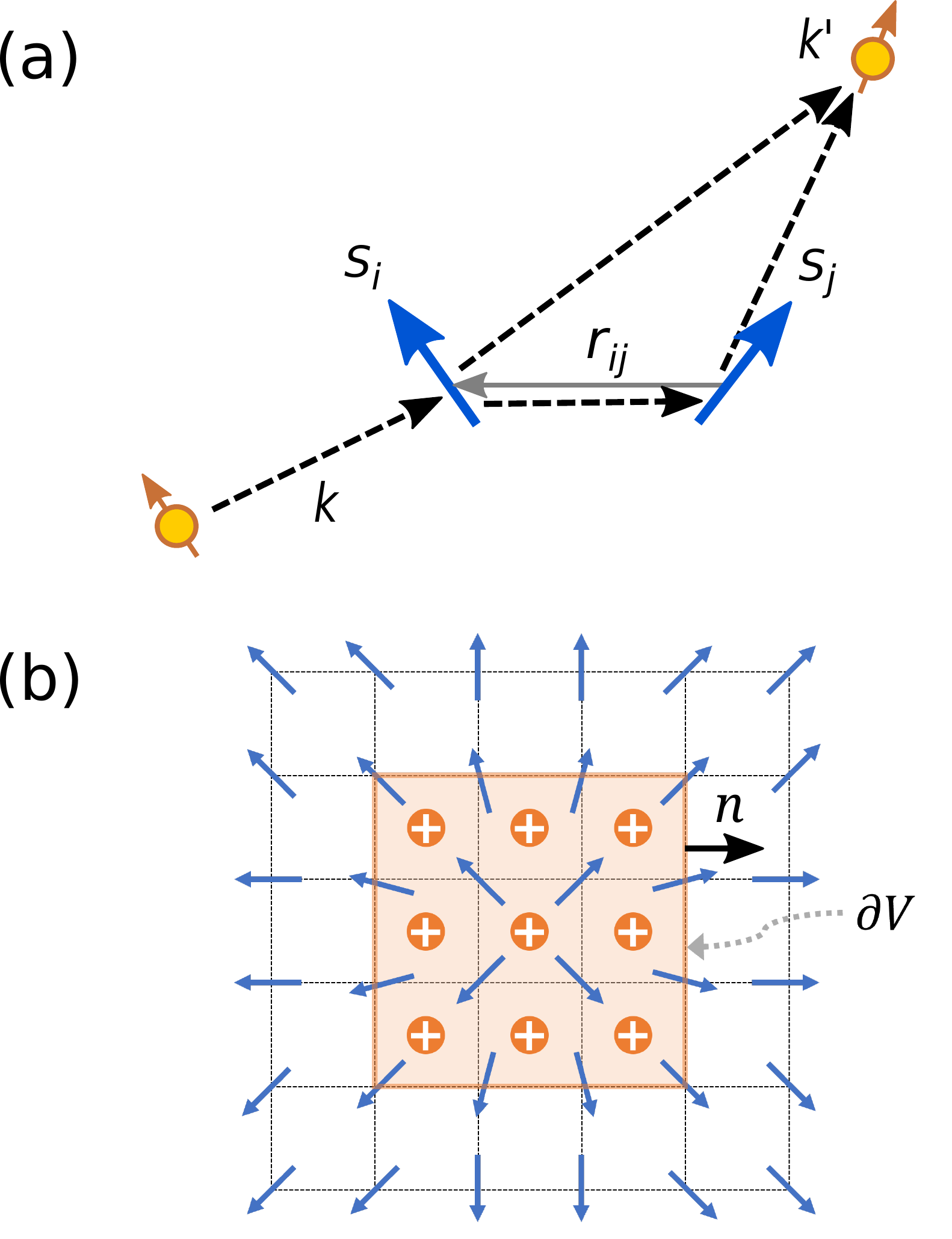}
  \caption{
  Schematic of magnetic monopoles and skew scattering.
  {\bf a}. A schematic of skew scattering involving two spins. Here, $\bm r_{ij}$ is the vector connecting two spins.
  {\bf b}. An example of magnetic monopole. $V$ and $\partial V$ denotes the area and its surface, respectively. $\bm n$ is the normal unit vector on $\partial V$.
  }\label{fig:model}
\end{figure}

%%%%%%%%%%%%%%%%%%%%%%%%%

%\section{Results}
%\subsection{Rashba Hamiltonian coupled to $XY$ spins}
As a demonstration, we first focus on the Rashba electrons coupled to localized classical $XY$ spins by Kondo coupling.
The Hamiltonian reads
\begin{align}
H=&H_0+H_K,\label{eq:HXY}\\
H_0=&\sum_{\bm k,\alpha,\beta}c_{\bm k\alpha}^\dagger\left[\left(\frac{\hbar^2k^2}{2m}-\mu\right)\delta_{\alpha\beta}+\hbar\lambda\hat z\cdot\bm k\times\bm\sigma_{\alpha\beta}\right]c_{\bm k\beta},\nonumber\\
H_K=&-Ja^2\sum_{i,\alpha,\beta}c_{\alpha}^\dagger(\bm r_i)[\bm S(\bm r_i)\cdot\bm\sigma_{\alpha\beta}]c_\beta(\bm r_i),\nonumber
\end{align}
where $c_{\bm k\alpha}$ ($c_{\bm k\alpha}^\dagger$) is the annihilation (creation) operator of the electron with momentum $\bm k=(k_x,k_y)$ and spin $\alpha$, $c_{\alpha}(\bm r)=\frac1V\sum_{\bm k}e^{{\rm i}\bm k\cdot\bm r}c_{\bm k\alpha}$ [$c_{\alpha}^\dagger(\bm r)=\frac1V\sum_{\bm k}e^{-{\rm i}\bm k\cdot\bm r}c_{\bm k\alpha}^\dagger$] is the annihilation (creation) operator of the electron at $\bm r$ and spin $\alpha$, $\bm S(\bm r_i)=(S^x(\bm r_i),S^y(\bm r_i))$ is the classical $XY$ spin at $\bm r_i$, and  $\bm\sigma=(\sigma^x,\sigma^y,\sigma^z)$ is the vector of Pauli matrices $\sigma^{x,y,z}$. The constant $m$ is the effective mass of the electron, $\mu$ is the chemical potential, $\lambda$ is the spin-orbit coupling, $J$ is the Kondo coupling, $a$ is the lattice constant, and $\hbar$ is the Planck constant.
The triple and inner products are $\hat z\cdot\bm k\times\bm\sigma=k_x\sigma^y-k_y\sigma^x$ and $\bm S(\bm r)\cdot\bm\sigma=S^x(\bm r)\sigma^x+S^y(\bm r)\sigma^y$.
The eigenenergy of $H_0$ reads $\varepsilon_{\bm k\eta}=\frac{\hbar^2k^2}{2m}+\eta\hbar\lambda k$ with $\eta=\pm1$.

We investigate the Hall effect by evaluating the skew scattering probability using a Born approximation and calculating the Hall conductivity using the scattering probability [See Method for details].
The scattering probability from $\bm k\eta$ state to $\bm k'\eta'$ is calculated within second-Born approximation, in which the skew scattering probability $W^-_{\bm k\eta\to\bm k'\eta'}=(W_{\bm k\eta\to\bm k'\eta'}-W_{\bm k'\eta'\to\bm k\eta})/2$ reads 
\begin{align}
    W^-_{\bm k\eta\to\bm k'\eta'}&=\frac{\pi W_1}{(aL^2)^2}\delta(\varepsilon_{k\eta}-\varepsilon_{k'\eta'})\notag\\
    &\qquad\times\sum_{i,j,l}(\bm S_l\times\bm r_{ij})\cdot(\bm S_i\times\bm S_j)\sin\delta\phi,
\label{eq:Wskew}
\end{align}
to the leading order in $ka$, where $L$ is the length of the system, $\delta\phi={\rm asin}(\hat{z}\cdot \frac{\bm k\times\bm k'}{kk'})$ is the scattering angle
, and $\bm{S}_i=\bm{S}(\bm{r}_i)$ is the magnetic moment at $\bm{r}_i$, .
The coefficient $W_{1}$ is shown in Tab.~\ref{tab:skew} in the Method section.
The terms in Eq.~\eqref{eq:Wskew} is of the order of $(ka)^2$, which is a lower order than the scalar-spin-chirality term in ${\cal O}\left((ka)^3\right)$~\cite{Ishizuka2018a}.
Hence, these terms potentially gives a larger contribution to the AHE than the scalar-spin-chirality ones, especially when the Fermi surface is small.
Moreover, the scattering term in Eq.~\eqref{eq:Wskew} induce a skew scattering without a finite scalar spin chirality nor magnetic moment perpendicular to the plane, both of which is zero in the $XY$ models.

We calculate the Hall conductivity by the skew scattering using the semiclassical Boltzmann theory [See Method for details].
The conductivity using Eq.~\eqref{eq:Wskew} reads
\begin{align}
\sigma_{xy}=-\frac{\sigma_1}{aL^2}\sum_{i,j,l}(\bm S_l\times\bm r_{ij})\cdot(\bm S_i\times\bm S_j).\label{eq:sxy2d}
\end{align}
Here, the coefficient $\sigma_{1}$ is in Tab.~\ref{tab:sxy}, where we introduce a phenomenological relaxation rate $\tau$.
The coefficients shows different $\mu$ dependence reflecting the distinct nature of electron bands above and below $\mu=0$.
A similar result for Heisenberg spin is given in Supplemental Information.

The outer product in the sum of Eq.~\eqref{eq:sxy2d} shows that the scattering by two spins are necessary for a finite $\sigma_{xy}$ in this system.
Note that, due to the zero out-of-plane magnetization, the AHE contributions known in ferromagnets vanishes.
However, a finite $\sigma_{xy}$ appears due to the scattering by multiple spins.

%%%%%%%%%%%%%%%%%%%%%%%%%
\begin{table}
  \begin{tabular}{c|c|c}
  \hline
  & $\mu<0$ & $0\le\mu$ \\
  \hline
  $\sigma_1$ & $\displaystyle{\frac{e^2\tau^2}{2\pi^2}\frac{J^3a^5}{\hbar^8}\frac{m^4\lambda^2a(\mu+m\lambda^2)}{\sqrt{(\lambda m)^2+2m\mu}}}$ & $\displaystyle{\frac{e^2\tau^2}{2\pi^2}\frac{J^3a^5}{\hbar^8}m^3\lambda a(\mu+m\lambda^2)}$ \\
  \hline
  $\sigma_2$ & $\displaystyle{\frac{e^2\tau^2}{4\pi^2}\frac{J^3a^5}{\hbar^7}\frac{m^4\lambda^3}{\sqrt{(\lambda m)^2+2m\mu}}}$ & $\displaystyle{\frac{e^2\tau^2}{4\pi^2}\frac{J^3a^5}{\hbar^7}m^3\lambda^2}$ \\
  \hline
  $\sigma_3$ & $\displaystyle{\frac{e^2\tau^2}{4\pi^2}\frac{J^3a^5}{\hbar^8}\frac{m^4\lambda^2a(\mu+m\lambda^2)}{\sqrt{(\lambda m)^2+2m\mu}}}$ & $\displaystyle{\frac{e^2\tau^2}{4\pi^2}\frac{J^3a^5}{\hbar^8}m^4\lambda^3a}$ \\
  \hline
  \end{tabular}
  \caption{
  Coefficients of the Hall conductivities in Eq.~\eqref{eq:sxy2d}.
  The center column is for chemical potential $\mu<0$ and the right column is for $\mu\ge0$.
  }\label{tab:sxy}
\end{table}
%%%%%%%%%%%%%%%%%%%%%%%%%%%%%%%%

%\subsection{Skew scattering by magnetic monopoles}
Focusing on the scattering processes involving two nearest-neighbor spins, $i$ and $j$, the Hall conductivity reads
\begin{align}
    \sigma_{xy}=\frac{2\sigma_1}{aL^2}\sum_{\langle i,j\rangle}(1+\bm S_i\cdot\bm S_j)[\bm r_{ij}\cdot(\bm S_i-\bm S_j)].\label{eq:sxy2d_cont}
\end{align}
In the continuum limit assuming $\bm S_i\sim\bm S_j$, this formula becomes
\begin{align}
    \sigma_{xy}=\frac{4\sigma_1}{aL^2}\int_V dx^2\;\nabla\cdot\bm S(\bm x),\label{eq:sxy2d_cont2}
\end{align}
where $\nabla=(\partial_x,\partial_y)$ and $\nabla\cdot\bm S=\partial_x S^x(\bm r)+\partial_y S^y(\bm r)$.
Here, we used the gradient expansion and $|\bm S_i|=1$.
The divergence $\nabla\cdot\bm S(\bm x)$ defines the magnetic monopole charge of spins, analogous to the definition of electric charge in electromagnetism.

Using the divergence theorem, the above formula reads
\begin{align}
    \sigma_{xy}=\frac{4\sigma_1}{aL^2}\int_{\partial V}\bm n(\bm x)\cdot\bm S(\bm x)dl,\label{eq:gauss}
\end{align}
as in the Gauss' theorem in electromagnetism.
Here, the integral is taken over the boundary of the system $\partial V$, where $\bm n(x)$ is the unit vector perpendicular to the boundary [Fig.~\ref{fig:model}(b)].
Hence, the anomalous Hall conductivity is related to the net magnetic charge, not to the fine details of the spin structure.

%\subsection{Anomalous Hall effect}
As an example of the magnetic monopole, we first discuss the vortex in Fig.~\ref{fig:model}(b).
The spin configuration reads $\bm S(\bm r)=(\cos(\phi),\sin(\phi),0)$, where $\phi$ is the azimuth. The Hall conductivity reads
\begin{align}
    \sigma_{xy}=\frac{8\pi\sigma_1}{aL}.
\end{align}
A finite Hall effect by the magnetic monopole implies that the monopoles are electrically detectable using the Hall effect, similar to the detection of skyrmions using anomalous Hall effect.

%%%%%%%%%%%%%%%%%%%%%%%%%%
\begin{figure}
\includegraphics[width=\linewidth]{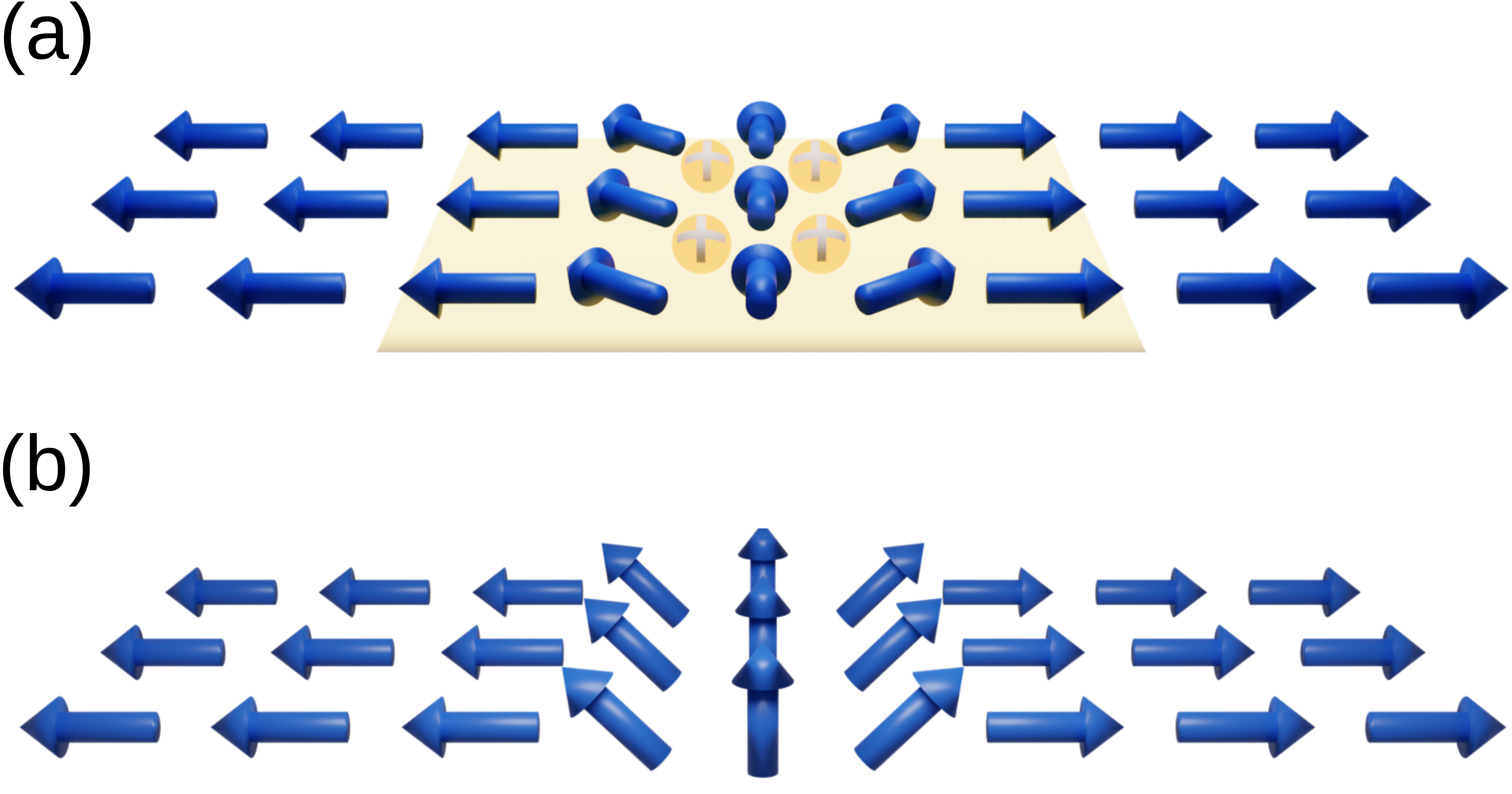}
  \caption{
  Schematics of domain walls. (a) A domain wall whose spin rotates in the $xy$ plane. In view of magnetic monopoles, the domain wall corresponds to magnetic monopoles (yellow circles) aligned along the domain wall. (b) A similar argument holds for domain walls with spins rotating out-of-plane.
  }\label{fig:domainwall}
\end{figure}
%%%%%%%%%%%%%%%%%%%%%%%%%%%%%%%%%%%%%%%%%%%%%%%%%%%

A more familiar example of the magnetic monopole is the domain walls of ferromagnets [Fig.~\ref{fig:domainwall}].
The domain wall is a boundary between the two ferromagnetic domains whose magnetization points along the opposite directions [Fig.~\ref{fig:domainwall}].
For the $XY$ spins, the spins rotate within the $xy$ plane [Fig.~\ref{fig:domainwall}(a)].
In view of magnetic monopole, this domain wall corresponds to a line of magnetic charges.
Mathematically, the correspondence is explicitly shown by applying the Gauss' law to the region surrounding the wall [yellow region in Fig.~\ref{fig:domainwall}(a)].
Note that the $y$ component of spins does not contributes to the monopole charge because it is uniform along the $y$ axis.
Using Eq.~\eqref{eq:gauss}, the Hall conductivity reads
\begin{align}
    \sigma_{xy}=\frac{8\sigma_1}{aL}.
\end{align}
Note that the conductivity scales $L^{-1}$ to the system size $L$.
However, it gives an observable consequence if $\sigma_1$ is sufficiently large.

To examine the typical magnitude of the response, we estimate the Hall conductivity using a set of values for Rashba electron known in the experiment~\cite{Ishizaka2011}: $m=10^{-30}$ kg, $\mu=0.19$ eV, $a=4.3$ \AA, and $\hbar\lambda=3.8$ eV\AA.
Assuming the Kondo coupling $J=0.1$ eV, the Hall conductivity becomes $\sigma_{xy}\sim10^4\;\Omega^{-1}$cm$^{-1}$ for a domain wall in $L=1\;\mu$m in size device.
The number is sufficiently large for the observation in experiments~\cite{Nagaosa2010}.
The Hall conductivity is determined solely by the direction of magnetic moments in the two adjacent domains or the magnetization at the boundary of the device [the boundary of the shaded region in Fig.~\ref{fig:domainwall}(a)], as discussed above.
In other words, the detail of the domain wall structure does not affect the result. 
The robustness against the fine structure of domain wall implies that this AHE is potentially a good probe of domain walls.

%%%%%%%%%%%%%%%%%%%%%%%%%%%%%%%%%%%%%%%%%%%%%%%%%%%
\begin{table}[tb]
    \begin{tabular}{c|cccccccccc}\hline
        &$\pi$ &$C^z_2$&$C^x_2$ &$C^y_2$ &$M_z$ &$M_x$ &$M_y$ &$C^z_3$ &$C^z_4$ &$C^z_6$\\ \hline
       $\nabla\cdot\bm S_\perp$ & 0 & - &0 &0 &0 &- &- &- &- &- \\
       $(\nabla\times\bm S_\perp)^z$ & 0 &- &- &- &0 &0 &0 &- &0 &- \\
       $\sigma^1_\mathrm{Dressel}$ & 0 &- &- &- &0 &0 &0 &0 &- &0 \\
       $\sigma^2_\mathrm{Dressel}$ & 0 &- &- &- &0 &0 &0 &- &- &- \\ \hline
       Rashba & $\times $ &$\circ$ &$\times$ &$\times$ &$\times$ &$\circ$ &$\circ$ &$\circ$ &$\circ$ &$\circ$ \\
       Dresselhaus &$\times$ &$\circ$ &$\circ$ &$\circ$ &$\times$ &$\times$ &$\times$ &$\times$ &$\times$ &$\times$ \\ \hline
    \end{tabular}
    \caption{Symmetries of AHE and SOI. In the table, it is written as "0" if each conductivity in the phenomenological AHE formula vanishes under each symmetry.
    $\nabla\cdot\bm S_\perp$ and $(\nabla\times\bm S_\perp)^z$ are the Hall current proportional to the divergence and rotation of the magnetic texture, respectively.
    $\sigma^1_\mathrm{Dressel}$ and $\,\sigma^2_\mathrm{Dressel}$ are respectively the first and second terms in Eq.~\eqref{eq:DresselhausHall}.
    }
    \label{tab:sym}
\end{table}
%%%%%%%%%%%%%%%%%%%%%%%%%%%%%%%%%%%%%%%%%%%%%%%%%%%

%\subsection{Anomalous Hall effect and crystal symmetry}
The reason we find $\nabla\cdot\bm S$ term in the Rashba model can be understood from the symmetry viewpoint.
The AHE by magnetic monopole is phenomenologically described by a formula $J_y=\sigma\int\nabla\cdot\bm S_\perp(\bm r) dr^2\, E_x$.
For example, when the inversion symmetry exists in the system, the symmetry operation transforms $\bm S_\perp(\bm r)\!\!\to\!\bm S_\perp(-\bm r)$, $J_y\!\!\to\!-J_y$, and  $E_x\!\!\to\!-E_x$.
Hence, the phenomenological formula becomes $J_y=-\sigma\int\nabla\cdot\bm S_\perp(\bm r) dr^2\, E_x$.
Thus, $\sigma=-\sigma$ for the system with inversion symmetry, implying that the AHE proportional to $\nabla\cdot\bm S_\perp$ vanishes.
Similarly, the mirror operation about the $z$ axis, $M_z$: $x,y\to x,y$ and $z\to -z$, gives $\sigma=-\sigma$, again implying the Hall effect is prohibited.
The top half of Tab.~\ref{tab:sym} shows the symmetry requirements, where $0$ denotes the vanishing Hall conductivity.
The lower half of the table shows the symmetry of Rashba and Dresselhaus models, where the $\times$ in the table denotes the symmetry do not exist in the model. 

Reflecting the symmetry property, the $\nabla\cdot\bm S$ term is allowed in Rashba model whereas the Hall effect related to other form of spin structure is possible in the Dresselhaus model.
Applying the same method to Dresselhaus model, we find that the Hall conductivity reads
\begin{align}
    \sigma_{xy}&=\frac{2\sigma_1}{aL^2}\int dx^2\bigl[S^x(\bm{S}\times\partial_x\bm{S})_z-S^y(\bm{S}\times\partial_y\bm{S})_z\bigr]\nonumber \\
    &\quad -\frac{2\sigma_1}{aL^2}\int dx^2[\partial_x S^y+\partial_y S^x]
    \label{eq:DresselhausHall}
\end{align}
in the continuum limit [See Supplementary Information for details].
Here, we find vanishing the $\nabla\cdot\bm S$ term as expected from Tab.~\ref{tab:sym}.
As demonstrated by the two models, the form of spin textures reflected to AHE is understandable from the crystal symmetry.

%%%%%%%%%%%%%%%%%%%%%%%%%%%%%%%%%%%%%%%%%%%%%%%%%%%
\begin{figure}
    \centering
    \includegraphics[width=0.8\linewidth]{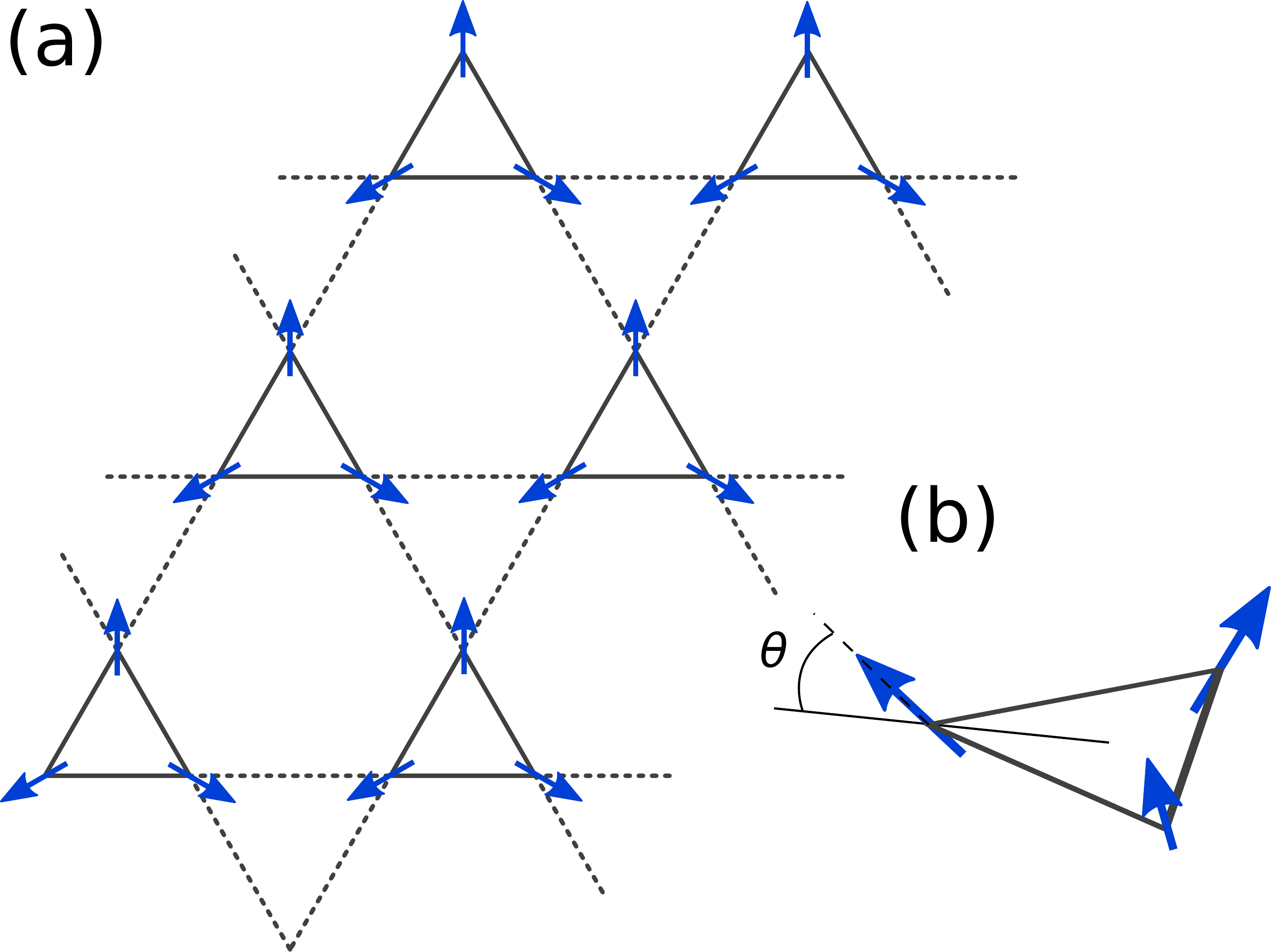}
    \caption{
    Schematic of (a) breathing kagom\'e lattice. The bond lengths for upward and downward triangle are different. (b) The out-of-plane canting angle is defined by the angle between the spin and the kagom\'e plane. 
    }\label{fig:kagome}
\end{figure}

%%%%%%%%%%%%%%%%%%%%%%%%%%%%%%%%%%%%%%%%%%%%%%%%%%%

%\subsection{Monopole-related Hall effect by lattice distortion}
We next look at the effect of the lattice distortion, in particular the formation of clusters.
In some materials, the lattice distortion forms a cluster of spins such as in trimerized triangular lattice~\cite{Uchida2021} and in breathing kagom\'e~\cite{Aidoudi2011,Clark2013} and pyrochlore~\cite{Okamoto2013,Ghosh2019} magnets. 
As a concrete example, we consider a breathing kagome lattice with canted 120$^\circ$ spin configuration as in Fig.~\ref{fig:kagome}(a).
By extending Eq.~\eqref{eq:sxy2d} to the spins with nonzero $S^z$, the Hall conductivity reads
\begin{align}
\sigma_{xy}=&12\sigma_1\delta\cos\theta(1+3\sin^2\theta)+24\sqrt3\sigma_2\sin^3\theta\nonumber\\
&-48\sigma_3\delta\cos\theta\sin^2\theta,
\end{align}
where $\theta$ is the out-of-plane canting angle, and the bond length of the upward (downward) triangle is $(\frac12+\delta)a$ [$(\frac12-\delta)a$] [Fig.~\ref{fig:kagome}(b)].
When $\theta=0$, i.e., for the coplanar 120$^\circ$ order, the conductivity becomes $\sigma_{xy}=12\sigma_1\delta$. 
Unlike the uniform case discussed in the previous sections, the Hall conductivity in the distorted kagom\'e lattice remains finite in the bulk limit.

In view of the magnetic monopole, the 120$^\circ$ order is a pair of monopole and anti-monopole (the monopole with the opposite charge).
Hence, the total monopole charge is zero which is consistent with the zero Hall conductivity at $\delta=0$.
In presence of the breathing, however, the breathing breaks the cancellation between monopole and anti-monopole contributions, giving a finite Hall effect.
The result shows that the monopole-related Hall effect also occurs in the bulk when the lattice distortion, such as breathing, exists in the lattice. 

%%%%%%%%%%%%%%%%%%%%%%%%%%%%%%%%%%%%%%%%%%%%%%%%%%%

%\section{Discussions}
In this work, we systematically study the skew scattering by multiple spins in systems with strong SOI.
Beside the skew scattering term similar to one discussed by Kondo~\cite{Kondo1962}, we find two terms contributing to the skew scattering.
These terms appear in the order of $(ka)^2$, in contrast to the $(ka)^3$ for the chirality-related AHE~\cite{Ishizuka2018a}.
Hence, the novel terms potentially give a larger contribution to the AHE in magnetic semiconductors where $ka$ is small due to the small Fermi surface, such as semiconductors described by Rashba and Dresselhaus models.
A particularly interesting term is that proportional to $(1+\bm S_i\cdot\bm S_j)[\bm r_{ij}\cdot(\bm S_i-\bm S_j)]$, which gives a skew scattering proportional to the magnetic monopoles in the continuous limit, $\nabla\cdot\bm S(\bm r)$.
We find that this is the only term that contributes to AHE when the spins lie in the $xy$ plane, e.g., for magnetic metals with the easy-plane anisotropy.
The $\nabla\cdot\bm S(\bm r)$ term gives rise to AHE in the presence of a vortex-like defect or domain walls, or in a material with breathing-type distortion.

The Hall effect proportional to $\nabla\cdot\bm S(\bm r)$ contributes to the AHE in the presence of a vortex-like structures in Fig.~\ref{fig:model}(b) and in domain walls [Fig.~\ref{fig:domainwall}(a)].
Importantly, the Hall conductivity depends only on the spin configuration of the bulk surrounding the vortex or the domain wall, and not to the fine details of the spin structure.
Such robustness to the detail is also interesting from application viewpoint, especially as the local probe for detecting magnetic structures which is a key technology in racetrack memory (Ref.~\cite{Fert2013,Tomasello2014,Maccariello2018}).

In the last, we note that a recent work studying the effect of SOI on the AHE by non-collinear magnetic states finds vanishing contribution from the SOI~\cite{Yokoyama2020}.
This work focuses the intrinsic AHE from the Berry phase viewpoint, in which they find only two relevant contributions: the conventional AHE by ferromagnetic moment and the chirality-related AHE.
In contrast, our study finds the skew scattering by magnetic moments is strongly affected by the SOI, which gives rise to rich features not seen in the systems without SOI.

\section{Method}
Here we provide the details of the scattering and Boltzmann theory. 
We investigate the scattering rate of the magnetic scattering processes  using a second Born approximation. 
The rate of scattering an electron with momentum $\bm{k}$ and spin $\alpha$ to that of $\bm{k}'$ and $\beta$ is given as   
\begin{align}
    W_{\bm k\alpha\rightarrow \bm k'\beta}
    =\frac{2\pi}{\hbar}\left|N^{(1)}_{\bm k\alpha,\bm k'\beta}+N^{(2)}_{\bm k\alpha,\bm k'\beta}\right|^2\,\delta(\varepsilon_{\bm k\alpha}-\varepsilon_{\bm k'\beta}),
    \label{eq:born}
\end{align}
where $N^{(1)}_{\bm k\alpha,\bm k'\beta}$ and $N^{(2)}_{\bm k\alpha,\bm k'\beta}$ are the first and second-order scattering amplitudes respectively. 
In our work, we treate the Kondo Hamiltonian $H_K$ as a perturbation to the non-perturbative Hamiltonian $H_0$ in Eq.~\eqref{eq:HXY}.
Then, scattering amplitudes are given by
\begin{align}
    N^{(1)}_{\bm k\alpha,\bm k'\beta}&=\langle\bm k \alpha|H_K|\bm k' \beta\rangle,\\
    N^{(2)}_{\bm k\alpha,\bm k'\beta}&=\sum_{\bm p ,\eta}\frac{\langle\bm k \alpha|H_K|\bm p\eta\rangle\langle\bm p\eta|H_K|\bm k' \beta\rangle}{\varepsilon_{\bm k\alpha}-\varepsilon_{\bm p \eta}-i0},
\end{align} 
where $|\bm k\alpha\rangle$ denotes an eigenstate of $H_0$ and $\varepsilon_{\bm k\alpha}$ is an eigenenergy of that. 
Since the band structure of the system with the SOI is qualitatively different between $\varepsilon_{\bm k\alpha}>0$ and $\varepsilon_{\bm k\alpha}<0$, physical properties such as skew scattering and transport coefficients change qualitatively through the sign of the chemical potential.
As the scattering process is elastic, $\varepsilon_{\bm{k}\alpha}=\varepsilon_{\bm{k'}\beta}$, we can express the magnitude of the momentum $k$ as a function of the chemical potential $\mu$ and the band index $\eta$,
\begin{align}
    k_\xi(\mu,\eta)=\frac{-\eta\lambda m+\xi\sqrt{\lambda^2m^2+2m\mu}}{\hbar},
\end{align}
where $\xi,\eta=\pm1.$
For $\mu\geq0$, only $\xi=+1$ appears while only $\eta=-1$ appears in $\mu<0$. 
To avoid redundancy, the notation as $k_\xi=k_\xi(\mu<0,\eta=-1)$, $k=k_{\xi=+1}(\mu\geq 0,\eta)$ and $k'=k_{\xi=+1}(\mu\geq0,\eta')$ are used in this section.

%%%%%%%%%%%%%%%%%%%%%%%%%
\begin{table}[tb]
  \begin{tabular}{c|c|c}
  \hline
  & $\mu<0$ & $0\le\mu$ \\
  \hline
  $W_1$ & $\displaystyle{(k_{\xi'}+k_{\xi})\frac{J^3a^5}{2\hbar^3}\frac{m^2\lambda}{\sqrt{m^2\lambda^2+2m\mu}}}$ & $\displaystyle{-(\eta k'+\eta'k)\frac{J^3a^5}{2\hbar^3}m}$ \\
    \hline
  $W_2$ & $\displaystyle{-\frac{J^3a^5}{\hbar^3}\frac{m^2\lambda}{\sqrt{m^2\lambda^2+2m\mu}}}$ & $\displaystyle{-\eta\eta'\frac{J^3a^5}{\hbar^3}m}$ \\
    \hline
  $W_3$ & $\displaystyle{\frac{J^3a^5}{2\hbar^4}\frac{m^3(\mu+m\lambda^2)}{\sqrt{m^2\lambda^2+2m\mu}}}$ & $\displaystyle{\eta\eta'\frac{J^3a^5\lambda}{2\hbar^4}m^2}$ \\
    \hline
  $W_4$ & $\displaystyle{(k_{\xi'}-k_{\xi})\frac{J^3a^5}{2\hbar^3}\frac{m^2\lambda}{\sqrt{m^2\lambda^2+2m\mu}}}$ & $\displaystyle{(\eta'k-\eta k')\frac{J^3a^5}{2\hbar^3}m}$ \\
    \hline
  \end{tabular}
  \caption{
  Coefficients of the skew scattering in the Rashba electron in Eqs.~\eqref{eq:Wskew} and \eqref{eq:Wskew2}.
  The left column is for chemical potential $\mu<0$ and the right column is for $\mu\ge0$.
  }\label{tab:skew}
\end{table}
%%%%%%%%%%%%%%%%%%%%%%%%%%%%%%%%

To examine the effect of the magentic scattering on the AHE, we mainly focus on the antisymmetric part of the scattering rate defined as
\begin{align}
    &W^{-}_{\bm k\alpha\rightarrow \bm k'\beta}=\frac{W_{\bm k\alpha\rightarrow \bm k'\beta}-W_{\bm k'\beta\rightarrow \bm k'\alpha}}{2} \nonumber\\
    &=\frac{4\pi^2}{\hbar}\sum_{\bm p ,\eta}\Im\bigl[\langle\bm k'\beta|H_K|\bm k\alpha\rangle\langle\bm k\alpha|H_K|\bm p\eta\rangle\langle\bm p\eta|H_K|\bm k'\beta\rangle\bigr] \nonumber\\
    &\qquad \times\delta(\varepsilon_{\bm k\alpha}-\varepsilon_{\bm k'\beta})\delta(\varepsilon_{\bm p\eta}-\varepsilon_{\bm k\alpha})\delta(\varepsilon_{\bm p\eta}-\varepsilon_{\bm k'\beta}),
    \label{eq:born-skew}
\end{align}
where $\delta(x)$ is the Dirac delta function.
We derived the second line of Eq.~\eqref{eq:born-skew} from Eq.~\eqref{eq:born}.
Equation~\eqref{eq:born-skew} for the Rashba system reads
\begin{align}
    &W^{-}_{\bm k\eta\rightarrow \bm k'\eta'}=\pi\delta(\varepsilon_{k\eta}-\varepsilon_{k'\eta'})\notag\\
    &\cdot\sum_{i,j,l}\biggl[\frac{W_1}{(aL^2)^2}(\bm S_l\times\bm r_{ij})\cdot(\bm S_i\times\bm S_j)-\frac{W_2}{(L^2)^2}S^z_l(\bm S_i\cdot\bm S_j)\nonumber \\
    &\qquad+\frac{W_3}{(aL^2)^2}S^z_l(\hat{z}\times\bm r_{ij})\cdot(\bm S_i\times\bm S_j)\biggr]\sin\delta\phi \nonumber\\
    &+\delta(\varepsilon_{k\eta}-\varepsilon_{k'\eta'})\frac{W_4}{(aL^2)^2}\sum_{i,j,l}(\bm S_i\cdot\bm S_j)(\hat{z}\cdot\bm r_{il}\times\bm S_l)\cos\delta\phi,
    \label{eq:Wskew2}
\end{align}
for Heisenberg spins $\bm S_i$; the $S_i^z=0$ case corresponds to the $XY$ spin case.  
Here, we neglect higher order terms of ${\cal O}\left((ka)^3\right), {\cal O}(J^4)$, and assume $2m\lambda/\hbar$ is same order of $k$. The skew terms are taken averaged over the mean angle $\bar{\phi}=(\phi_{k}+\phi_{k'})/2$, where $\phi_k$ is the direction of the electron momentum $\bm k$.
Then, Eq.~\eqref{eq:Wskew2} only depends on the scattering angle $\delta\phi=\phi_k-\phi_{k'}$. 
Each coefficient $W_{i}$ is summarized in in Tab.~\ref{tab:skew}.
As mentioned in the main text, all terms in Eq.~\eqref{eq:Wskew2} are different from scalar-spin-chirality term that appears without the SOI systems on the order of ${\cal O}((ka)^3)$. 

Next, we evaluate the Hall conductivity which arises from the skew scattering in Eq.~\eqref{eq:Wskew2} within the semi-classical Boltzmann theory.
In the presence of the uniform static electric filed $\bm{E}$, the Boltzmann equation reads
\begin{align}
    e\bm E\cdot\bm v_{\bm{k}\eta}f'_0(\mu)=\frac{g_{\bm k\eta}}{\tau}-\frac{V}{(2\pi)^2}\sum_{\eta}\int d^2k W^-_{\bm k'\eta'\rightarrow\bm k\eta}g_{\bm k'\eta'}, 
    \label{eq:boltzmann}
\end{align}
where $\bm v_{\bm{k}\eta}=\nabla_k\varepsilon_{k\eta}/\hbar$ is the velocity of the electron in the momentum $\bm k$, and $f_0(\varepsilon)$ is the Fermi-Dirac distribution with its energy derivative $f_0'(\varepsilon)$. 
We assume that the electron distribution is expanded as $f_{\bm k\eta}=f_0(\varepsilon_{\bm k\eta})+g_{\bm k\eta}$ and the displacement from the equilibrium distribution $g_{\bm k\eta}$ is order $E$.
For the scattering terms in the right-hand side of Eq.~\eqref{eq:boltzmann}, the symmetric part of the scattering rate $W^+_{\bm{k}\eta\rightarrow\bm{k}'\eta'}=(W_{k\eta\to k'\eta'}+W_{k'\eta'\to k\eta})/2$ is taken into account the relaxation time approximation with relaxation time $\tau$.

Equation \eqref{eq:boltzmann} is analytically solvable for $W^-$~\cite{Ishizuka2018a}.
In order to evaluate the displacement $g_{\bm{k}\eta}$, we define a parameter $\bm{P}_\eta(\mu)$ related to the integral of $g_{\bm k\eta}$ with respect to the angle as
\begin{align}
    \bm{P}_\eta(\mu)=\int_0^{2\pi}d\phi_k\, \hat{k}\,g_{\bm k\eta},
\end{align}
where $\hat{k}$ is the unit vector along $\bm{k}$.
Using $\bm{P}_\eta$, we rewrite the Boltzmann equation as
\begin{align}
    g_\eta=e\tau \bm{E}\cdot\bm{v}_{\bm{k}}f'_0(\varepsilon)+\tau \tilde{W}^-\sum_{\eta'}V^{\eta\eta'}(\mu)\,\hat{z}\cdot\hat{k}\times\bm{P}_{\eta'}(\mu),
    \label{eq:boltzmann2}
\end{align}
where $\tilde{W}^-$ and $V^{\eta\eta'}$ are the coefficients from Eq.~\eqref{eq:Wskew2} as follows,
\begin{align}
    \tilde{W}^-=\frac{1}{4\pi v(\mu)}\frac{J^3a^6}{L^2\hbar^4}m,
\end{align}
\begin{widetext}
\begin{align}
    &V^{\eta\eta'}(\mu)\nonumber\\
    &=\begin{cases}
        \frac{\lambda k_{\eta'}}{v(\mu)}\left[-\frac{1}{2}(k_{\eta'}+ k_{\eta})\sum_{ijl}(\bm S_l\times\bm r_{ij})\cdot(\bm S_i\times\bm S_j)+\sum_{ijl}S^z_l(\bm S_i\cdot\bm S_j)-\frac{\mu+m\lambda^2}{\hbar\lambda}\sum_{ijl}S^z_l(\hat{z}\times\bm r_{ij})\cdot(\bm S_i\times\bm S_j)\right] \\ \hspace{145mm} (\mu<0), \\
        k'\left[\frac{1}{2}(\eta k'+\eta' k)\sum_{ijl}(\bm S_l\times\bm r_{ij})\cdot(\bm S_i\times\bm S_j)
        +\eta\eta'\sum_{ijl}S^z_l(\bm S_i\cdot\bm S_j)-\eta\eta'\frac{m\lambda}{\hbar}\sum_{ijl}S^z_l(\hat{z}\times\bm r_{ij})\cdot(\bm S_i\times\bm S_j)\right] \\ \hspace{145mm} (\mu\geq0),
    \end{cases}
\end{align}
\end{widetext}
where $v(\mu)=|\bm{v}_{\bm{k}}|$.
By multiplying $\hat{k}$ to the both sides of Eq.~\eqref{eq:boltzmann2} and integrating over $\phi_k$, we get the following self-consistent equation respct to $\bm{P}_\eta$:
\begin{align}
     \bm{P}_\eta=e\tau\pi v \bm{E}f'_0(\mu)+\pi\tau \tilde{W}^-\sum_{\eta}V^{\eta\eta'}\,\bm{P}_{\eta'}(\mu)\times\hat{z}.
     \label{eq:eqP}
\end{align}
Eq.~\eqref{eq:eqP} is a linear equation of $\bm{P}_\eta$ and can be solved analytically.
TO the leading order of $J^3$, $\bm{P}_\eta$ is expressed as
\begin{align}
    \bm{P}_\eta(\mu)=e\tau\pi v f'_0(\mu)\bigl(\bm{E}-\pi\tau\tilde{W}^-\hat{z}\times\bm{E}\,\sum_{\eta'}V^{\eta\eta'}\bigr).
    \label{eq:eqP2}
\end{align}
Finally, the Hall conductivity is calculated using the current formula
\begin{align}
    \bm{j}=\frac{-e}{(2\pi)^2}\sum_{\eta}\int d^2k\, \bm{v}_{\bm{k}\eta}g_{\bm{k}\eta},
\end{align}
by substituting Eq.~\eqref{eq:eqP2} for Eq.~\eqref{eq:boltzmann2}.
The results for the $XY$ spins are given in Eq.~\eqref{eq:sxy2d}, and for the Heisenberg spins in the Supplementary Information.

%%%%%%%%%%%%%%%%%%%%%%%%%%%%%%%%%%%%%%%%
\section*{acknowledgement}
We are thankful to Y. Niimi and K. Masuki for discussions.
This work is supported by JSPS KAKENHI (Grant Number JP19K14649).

%%%%%%%%%%%%%%%%%%%%%%%%%%%%%%%%%%%%%
%%%%%%%%%%%%%%%%%%%%%%%%%%%%%%%%%%%%%

%
\end{document}

\preprint{APS/123-QED}

\title{
  Supplementary Information to \\
``{\it Skew scattering by magnetic monopoles and anomalous Hall effect in spin-orbit coupled systems}''
}

\author{Jun Mochida}
\affiliation{
Department of Physics, Tokyo Institute of Technology, Meguro, Tokyo, 152-8551, Japan
}
\affiliation{Department of Physics, University of Tokyo, 7-3-1 Hongo, Bunkyo-ku, Tokyo 113-0033, Japan
}
\author{Hiroaki Ishizuka}
\affiliation{
Department of Physics, Tokyo Institute of Technology, Meguro, Tokyo, 152-8551, Japan
}

\date{\today}

% \begin{abstract}
% \end{abstract}

\pacs{
}% PACS, the Physics and Astronomy
% Classification Scheme.

\maketitle
\onecolumngrid
%%%%%%%%%%%%%%%%%%%%%%%%%%%%%%%%%%%%%%%%%%%%%
%%%%%%%%%%%%%%%%%%%%%%%%%%%%%%%%%%%%%%%%%%%%%
\section{Skew scattering by Heisenberg spins}

%\subsection{Anomalous Hall effect of Heisenberg spins}
For the case of Heisenberg spins $\bm S_i=(S_i^x,S_i^y,S_i^z)$, whose magnetic moment can point perpendicular to the plane, the Hall conductivity for Rashba model reads
\begin{align}
&\sigma_{xy}=-\frac{\sigma_1}{aL^2}\sum_{i,j,l}(\bm S_l\times\bm r_{ij})\cdot(\bm S_i\times\bm S_j)
+\frac{\sigma_2}{L^2}\sum_{i,j,l}S_l^z(\bm S_i\cdot\bm S_j)-\frac{\sigma_3}{aL^2}\sum_{i,j,l}S_l^z(\hat z\times\bm r_{ij})\cdot(\bm S_i\times\bm S_j).\label{eq:sxy3d}
\end{align}
In addition to the $\sigma_1$ term in Eq.~(3), two additional terms ($\sigma_2$ and $\sigma_3$; the explicit form given in Tab.~I in the main text) appears reflecting the uniaxial anisotropy of Rashba model.
This formula reduces to Eq.~(3) when $S_i^z=0$.
The $\sigma_2$ term in Eq.~\eqref{eq:sxy3d} is the generalization of an scattering mechanism pointed out by Kondo~\cite{Kondo1962S}, in which they consider the scattering by single spin. 
On the other hand, $\sigma_3$ is another term related to the vector spin chirality of $\bm S_i$ and $\bm S_j$ that does not exist in a system without SOI.
All these terms appear in the $(ka)^2$ order in the expansion with respect to $ka$.
Note that the scalar spin chirality term discussed in the previous works~\cite{Ishizuka2018aS,Ishizuka2021S} appears in ${\cal O}((ka)^3)$, an order higher than the above terms.

The SOI-induced terms contribute to the AHE by isolated skyrmions~\cite{Fujishiro2021S} and other non-colinear magnetic states.
For a smooth magnetic structure on the square lattice, Eq.~\eqref{eq:sxy3d} reads
\begin{align}
&\sigma_{xy}=\frac{16\sigma_2}{a^2L^2}\int dx^2S^z+\frac{4\sigma_1}{aL^2}\int dx^2\nabla\cdot\bm S
+\frac{4\sigma_2}{aL^2}\int dx^2\left[\partial_xS^z+\partial_yS^z+S^z\left\{\bm S\cdot\partial_x\bm S+\bm S\cdot\partial_y\bm S\right\}\right]\nonumber\\
&\hspace{2cm}+\frac{4\sigma_3}{aL^2}\int dx^2S^z\left[(\bm S\times\partial_x\bm S)_y-(\bm S\times\partial_y\bm S)_x\right].\label{eq:sxy3d2}
\end{align}
The formula gives the Hall conductivity
\begin{align}
\sigma_{xy}&=16\sigma_2m_z+\frac{4\pi\sigma_3}{a}\Omega(l/\xi)\xi\, n_{\mathrm{sk}},\\
\Omega(x)&=\left[\frac{\pi}{2}+\tan^{-1}\left\{\sinh(x)\right\}+\frac{\sinh(x)}{\cosh^2(x)}\right]
\end{align}
for a N\'{e}el-type skyrmion and 
\begin{align}\sigma_{xy}=16\sigma_2m_z,
\end{align}
for the Bloch-type skyrmion, where $m_z=\frac1{a^2L^2}\int dx^2 S^z$ is magnetization of the skyrmion, $\ell,\,\xi$ are the radius and domain wall length of the skyrmion respectively.
Note that the monopole term vanishes because the spins aligns uniformly at the boundary.
This value should be compared to the chirality-related AHE~\cite{Ishizuka2018aS}, $\sigma_{xy}^\text{ssc}=-e^2\tau^2\frac{2k_F^7}{3\pi^6}\frac{J^3ma^9}{V\hbar^5}n_{\text{sk}}$, where $n_{\text{sk}}=\frac1{L^2}\int\frac{d^2x}{4\pi}\bm S\cdot\partial_x\bm S\times\partial_y\bm S$ is the skyrmion density.

Recent study on the chirality-related AHE in skyrmion reports a noticeable correction arising from the Rashba SOI~\cite{Lux2020}, whereas another study focusing on the Berry phase effect discusses vanishing correction from the SOI~\cite{Yokoyama2020}.
For the Hall effect by skew scattering, the terms in the second row of Eq.~\eqref{eq:sxy3d} gives a finite correction proportional to the net magnetization, whereas the third term in Eq.~\eqref{eq:sxy3d} gives a finite correction only in N\'eel-type skyrmions.

The $\sigma_2$ and $\sigma_3$ also contributes to the AHE in domain walls when the magnetic moments at the wall points out of plane [Fig.~\ref{fig:domainwall}(b)].
In this case, the Hall conductivity reads 
\begin{align}
\sigma_{xy}=16\sigma_2m_z+\frac{8\sigma_1}{aL}+\frac{8\sigma_3}{aL}.
\end{align} 
In this case, the contribution from $\sigma_2$ and $\sigma_3$ becomes larger than $\sigma_1$ in a large device.

\section{Dresselhaus model}

We note that the symmetry plays a key role in the SOI-induced novel terms.
To compare the result of Rashba model to a model with different symmetry, we consider Dresselhaus model which lacks the mirror symmetries about $x$ and $y$ axes, $M_x$ and $M_y$, respectively.
The Hamiltonian reads
\begin{align}
H=&H_0+H_K,\\
H_0=&\sum_{\bm k,\alpha}c_{\bm k\alpha}^\dagger\left(\frac{\hbar^2k^2}{2m}-\mu\right)c_{\bm k\alpha}
+\frac\beta\hbar\sum_{\bm k,\alpha,\alpha'}c_{\bm k\alpha}^\dagger(k_x\sigma^x_{\alpha\alpha'}-k_y\sigma^y_{\alpha\alpha'})c_{\bm k\alpha'},\nonumber
\end{align}
and $H_K$ is the same as in Eq.~(1).
By using the same method, the general formula for AHE conductivity reads
\begin{align}
&\sigma_{xy}=-\frac{\sigma_1}{aL^2}\sum_{i,j,l}(\bm S_i\times\bm S_j)_z(S^x_l r^x_{il}-S^y_l r^y_{il})-\frac{\sigma_1}{aL^2}\sum_{i,j,l}(\bm S_i\cdot\bm S_j)(S^x_l r^y_{il}+S^y_l r^x_{il})
-\frac{\sigma_2}{L^2}\sum_{i,j,l}S_l^z(\bm S_i\cdot\bm S_j)\nonumber\\
&+\frac{\sigma_1}{aL^2}\sum_{i,j,l}S^z_l((\bm S_i\times\bm S_j)_x r^x_{il}-(\bm S_i\times\bm S_j)_y r^y_{il})+\frac{\sigma_3}{aL^2}\sum_{i,j,l}S^z_l((\bm S_i\times\bm S_j)_x r^x_{ij}-(\bm S_i\times\bm S_j)_y r^y_{ij}),\label{eq:sxy3d_Dressel}
\end{align}
where the coefficients $\sigma_i$ are given in Tab.~\ref{tab:sym}.
This formula reduces to
\begin{align}
\sigma_{xy}=&\frac{\sigma_1}{aL^2}\left[-\sum_{i,j,l}(\bm S_i\times\bm S_j)_z(S^x_l r^x_{il}-S^y_l r^y_{il})+\sum_{i,j,l}(\bm S_i\cdot\bm S_j)(S^x_l r^y_{il}+S^y_l r^x_{il})\right],
\label{eq:sxy2d_Dressel}
\end{align}
for the XY spin, i.e., $S_i^z=0$.
These two terms are different from the conductivity in the Rashba model.
The continuous limit of Eq.~\eqref{eq:sxy2d_Dressel} is calculated as follows
\begin{align}
    \sigma_{xy}&=\frac{2\sigma_1}{aL^2}\int dx^2\bigl[S^x(\bm{S}\times\partial_x\bm{S})_z-S^y(\bm{S}\times\partial_y\bm{S})_z\bigr] -\frac{2\sigma_1}{aL^2}\int dx^2[\partial_x S^y+\partial_y S^x].
\end{align}

The difference of $\sigma_{xy}$ between the Rashba and Dresselhaus models is understandable from the symmetry viewpoint. See Tab.~II in the main text.

%%%%%%%%%%%%%%%%%%%%%%%%%%%%%%%%%%%%%%%%%%%%%%%%%%%
\begin{table}[hb]
  \begin{tabular}{c|c|c}
  \hline
  & $\mu<0$ & $0\le\mu$ \\
  \hline
  $\sigma_1$ & $\displaystyle{\frac{e^2\tau^2}{4\pi^2}\frac{J^3a^5}{\hbar^{11}}\frac{m^4\beta^2a(\hbar^2\mu+m\beta^2)}{\sqrt{(\beta m)^2+2m\hbar^2\mu}}}$ & $\displaystyle{\frac{e^2\tau^2}{4\pi^2}\frac{J^3a^5}{\hbar^{11}}m^3\beta a(\hbar^2\mu+m\beta^2)}$ \\
  \hline
  $\sigma_2$ & $\displaystyle{\frac{e^2\tau^2}{4\pi^2}\frac{J^3a^5}{\hbar^9}\frac{m^4\beta^3}{\sqrt{(\beta m)^2+2m\hbar^2\mu}}}$ & $\displaystyle{\frac{e^2\tau^2}{4\pi^2}\frac{J
  ^3a^5}{\hbar^9}m^3\beta^2}$ \\
  \hline
  $\sigma_3$ & $\displaystyle{\frac{e^2\tau^2}{4\pi^2}\frac{J^3a^5}{\hbar^{11}}\frac{m^4\beta^2a(\hbar^2\mu+m\beta^2)}{\sqrt{(\beta m)^2+2m\hbar^2\mu}}}$ & $\displaystyle{\frac{e^2\tau^2}{4\pi^2}\frac{J^3a^5}{\hbar^{11}}m^4\beta^3a}$ \\
  \hline
  \end{tabular}
  \caption{
  Coefficients of the Hall conductivities in Eqs.~\eqref{eq:sxy3d_Dressel}.
  The center column is for chemical potential $\mu<0$ and the right column is for $\mu\ge0$.
  }\label{tab:sym}
\end{table}
%%%%%%%%%%%%%%%%%%%%%%%%%%%%%%%%%%%%%%%%%%%%%%%%%%%

%%%%%%%%%%%%%%%%%%
\newcommand{\bibitemS}[1]{
\stepcounter{count}
\bibitem[S\thecount]{#1}}
\newcounter{count}
%%%%%%%%%%%%%%%%%%